\documentclass[11pt]{article}
\usepackage{moriond,epsfig,psfrag}

\def\Journal#1#2#3#4{{#1} {\bf #2}, #3 (#4)}

\def\PRD{{\em Phys. Rev.} D}

\def\be{\begin{equation}}
\def\ee{\end{equation}}
\def\bea{\begin{eqnarray}}
\def\eea{\end{eqnarray}}

\def\beaa{\begin{array}}
\def\eeaa{\end{array}}
\newcommand{\nn}{\nonumber}
\begin{document}
\vspace*{4cm}
\title{Decays of MSSM Higgs in Flavour-Changing Quark Channels}

\author{ Ana~M.~Curiel and Mar\'{\i}a~J.~Herrero}

\address{Dept. de F\'{\i}sica Te\'orica, C-XI, Facultad de Ciencias, Universidad Aut\'onoma de Madrid, Cantoblanco, 28049 Madrid, Spain}

\maketitle\abstracts{We compute the genuine SUSY one-loop quantum contributions to flavour-changing MSSM Higgs-boson decays into $b \bar s$ and $s \bar b$ using the full diagrammatic approach that is valid for all $\tan \beta$ values and do not rely on the mass-insertion approximation for the characteristic flavour-changing parameter. We analyze in full detail the dependence of these flavour-changing partial widths on all the relevant MSSM parameters and also study the non-decoupling behaviour of these widths with the SUSY mass parameters. We find that these contributions are sizable as compared to the SM ones, and can be very efficient as an indirect method in the future search for Supersymmetry. }

\section{Introduction}
Rare processes involving Flavour Changing Neutral Currents (FCNC), and in particular, the ones related to Higgs physics, provide an 
extremely useful tool to investigate new physics beyond the Standard 
Model (SM). In the SM, there is a strong suppression of these FCNC due to the 
GIM-cancellation mechanism. In contrast, the Minimal Supersymmetric Standard Model (MSSM)
provides a natural framework 
where induced FC scalar interactions could be significant if the  
soft SUSY-breaking mass terms contain some 
non-diagonal structure in flavour space.

Here, we are going to concentrate on the neutral MSSM Higgs-boson decays into $b  \bar s$ and $s \bar b$ that are induced by SUSY loop contributions, in a general scenario that includes both possible flavour changing sources, via the Cabibbo-Kobayashi-Maskawa (CKM) matrix and via the misalignment between the quark and squark sectors. We find that these FCNC effects are large, considerably enhanced with respect to the SM contribution, and therefore, they are sensitive to search for indirect SUSY signals. 
  We also checked that our results are perfectly in agreement with the 
ones of the effective--Lagrangian approach in the large $\tan \beta$ and 
small FC parameter $\lambda$ limit.

One of the most important features of these observables are the remaining non-vanishing contributions even in the limit of very heavy
SUSY particles and, in addition, that they are enhanced by large
$\tan \beta$ factors. Such a  non-decoupling behaviour of SUSY particles 
in the Flavour Changing Higgs Decays (FCHD) can be of special
interest for indirect SUSY searches at future colliders, as the forthcoming 
LHC and a next $e^+e^-$ linear collider, in particular if the
SUSY particles turn out to be too heavy to be produced directly.

 Here we performed an exact computation 
of the complete SUSY-QCD and SUSY-EW one-loop contributions from squark--gluino, squark--chargino and squark--neutralino loops to the flavour-nondiagonal 
decay rates of the three neutral MSSM Higgs bosons, and therefore, our computation is valid for all values of the
characteristic parameter measuring the squark-mixing strength $\lambda$ and for all 
$\tan \beta$ values. 

\section{Flavour-changing interactions in the MSSM}
In the MSSM there are two sources of FC phenomena. The first one 
is common
to the Standard Model case and is due to mixing in the quark sector. It is 
produced by 
the different rotation in the $d$- and $u$-quark sectors, and its strength 
is driven by the off-diagonal CKM-matrix elements. This mixing produces 
FC electroweak
interaction terms involving  charged currents and, in particular, in this case,  SUSY FC electroweak  
interaction terms of the chargino--quark--squark type. 
The second source of FC phenomena is due to the possible 
misalignment between the rotations that diagonalize the quark and 
squark sectors.  When the squark-mass matrix is
expressed in the basis where the squark fields are parallel to the physical
quarks (the super-CKM basis), it is in general
non-diagonal in flavour space. This quark--squark misalignment produces 
new FC terms in
neutral-current as well as in charged-current interactions. 

Assuming that the non-CKM
squark mixing is significant only for transitions between the 
third- and second-generation
squarks, and that there is only LL mixing, the squark squared-mass matrix that has to be diagonalized via a $4 \times 4$ matrix, $R^{(d)}$, in the 
($\tilde s_L$,$\tilde s_R$,$\tilde b_L$,$\tilde b_R$) basis (similarly for the ($\tilde c_L$,$\tilde c_R$,$\tilde t_L$,$\tilde t_R$) basis with rotation matrix $R^{(u)}$), can be written as follows,
 \begin{eqnarray} 
M^2_{\tilde d} =\left\lgroup 
         \beaa{llll}
          M_{L,s}^2  &   m_s X_s & \lambda_{LL} M_{L,s} M_{L,b}& 0\\
           m_s X_s   &  M_{R,s}^2  & 0  &  0\\
          \lambda_{LL} M_{L,s} M_{L,b} & 0 & M_{L,b}^2  &   m_b X_b\\
          0 & 0 &  m_b X_b &  M_{R,b}^2
\eeaa
         \right\rgroup
\label{eq.dsquarkmass}
\end{eqnarray}
where $M_{L,q}^2$, $M_{R,q}^2$ and $m_q X_{q}$ are the usual flavour preserving entries of the MSSM squark squared mass matrices. The previous assumption~(\ref{eq.dsquarkmass}) is supported on many mSUGRA inspired models.

In our parametrization
of flavour mixing in the squark sector, 
there is only one free parameter, $\lambda_{LL} \equiv \lambda$, that characterizes the
flavour-mixing strength. Obviously, the choice
 $\lambda=0$ represents the case of zero  flavour mixing via misalignment.

After performing the whole diagonalization, we computed the loop-induced flavour-changing neutral Higgs boson decays into second and third generation quarks whose detailed results can be found in~\cite{Maria,Maria2}. In the next section we will discuss the dependence of the decay rates and branching ratios on the MSSM parameters and $\lambda$.   

\section{Numerical analysis}

\psfrag{l}{$\lambda$}
\psfrag{Brh0}{$Br (h^0 \rightarrow b \bar s + s \bar b)$}
\psfrag{BrH0}{$Br (H^0 \rightarrow b \bar s + s \bar b)$}
\psfrag{GA0}{$\Gamma (A^0 \rightarrow b \bar s + s \bar b)$}
\psfrag{GHH}{$\Gamma (H^0 \rightarrow b \bar s + s \bar b)$}
\psfrag{Gh0}{$\Gamma (h^0 \rightarrow b \bar s + s \bar b)$}
\psfrag{mu}{$\,\mu$}
\psfrag{M0}{$\,M_0$}

\begin{figure}
\psfig{figure=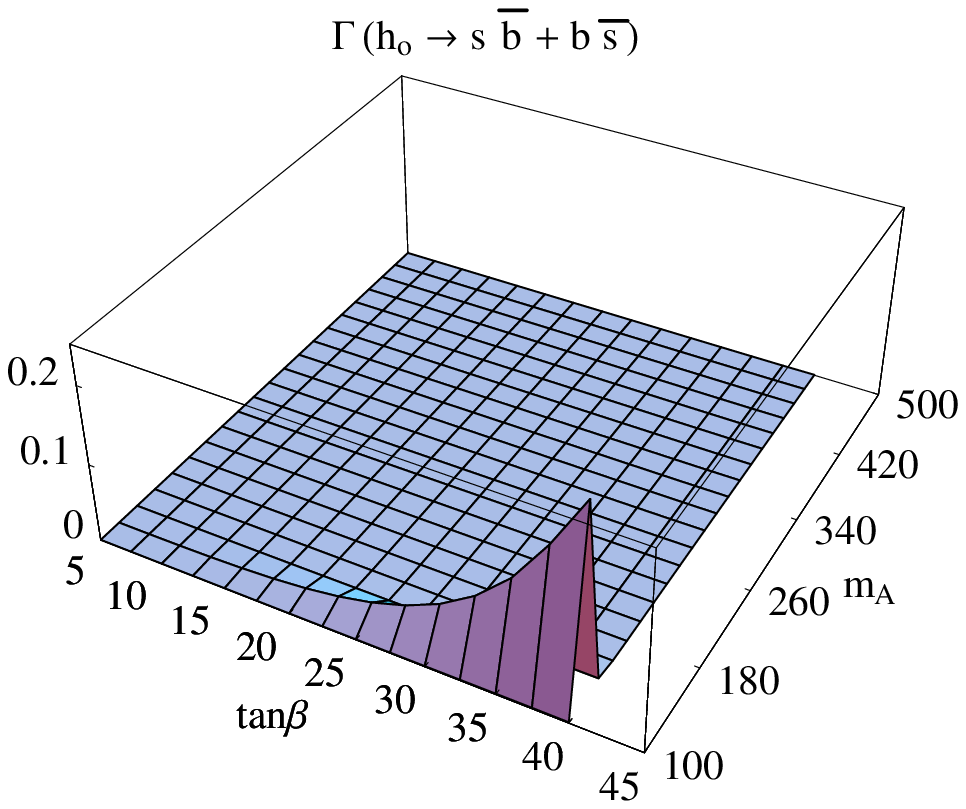,height=1.5in}
\psfig{figure=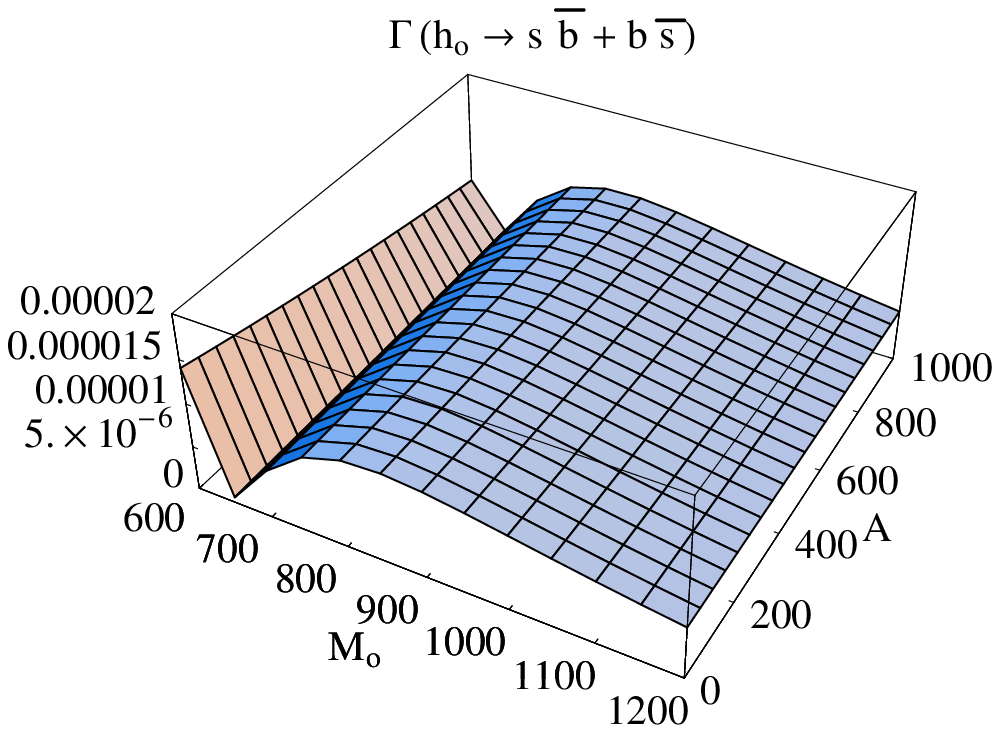,height=1.5in}
\psfig{figure=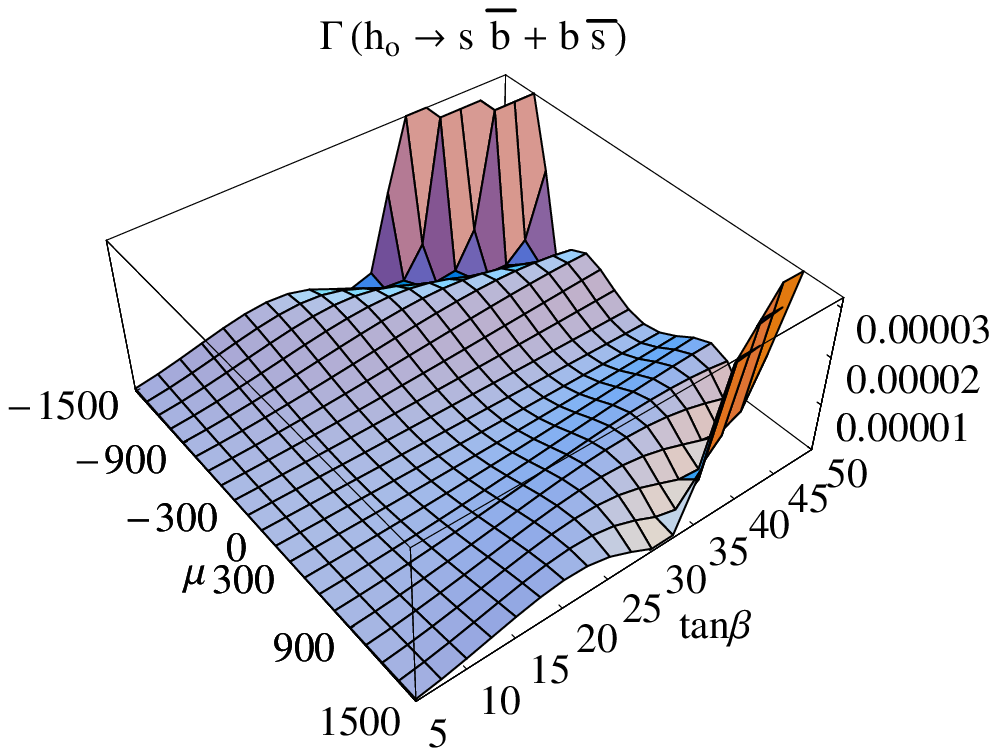,height=1.5in}
\caption{SUSY-QCD $\Gamma (h_0 \to b \bar s + s \bar b)$ (GeV) as a function of ($m_A$ (GeV), $\tan \beta$), ($M_0$ (GeV), A (GeV)) and ($\mu$ (GeV), $\tan \beta$). Fixed values chosen are $\mu=1500$ GeV, $M_0=600$ GeV, $M_{\tilde g}=300$ GeV, A=200 GeV, $m_A=250$ GeV, $\tan \beta=35$ and $\lambda=0.5$.
\label{contour}}
\end{figure}

Here we numerically estimate the size of the 
loop-induced FCHD as a function of the MSSM parameters and 
the mixing parameter $\lambda$. The GUT relations
$M_3 = \alpha_s/\alpha \, s_W^2\, M_2$ and $M_1=5/3 \, s_W^2/c_W^2 \, M_2$
are assumed. For the numerical analysis of the FCHD rates, 
only values of $\lambda$ (in the range $0\leq \lambda \leq 1$)
that lead to physical squark masses above 150 GeV 
will be considered. 
   
\psfrag{MA0}{$\,m_A$}
\psfrag{tb}{{$\tan \beta$}}

The MSSM parameters needed to determine the partial widths 
$\Gamma (H_x \to b \bar s + s \bar b)$, for $H_x \equiv h^0, H^0, A^0$, 
are the following six, 
$m_A$, $\tan \beta$, $\mu$, $M_{2}$, $M_0$, and $A$, where we have 
chosen, for simplicity, $M_0$ as a common value for the soft SUSY-breaking
squark mass parameters,
and all the various trilinear parameters, A, to be universal. 
 These parameters were varied over a broad range, and grouped into different pairs in order to visualize the individual dependences
of the FCHD widths for each neutral Higgs boson.
The most important thing to notice is that we obtain large values of the decay
widths. Besides, as expected, the SUSY-QCD contributions dominate by large the SUSY-EW ones, being at least one order of magnitude larger. On the other hand, a common  clear behaviour of all three decay widths is the increase 
with $\tan \beta$, yielding maximal FC effects
at large $\tan \beta$ values. In the following we comment shortly on the $h_0$ and $H_0$ decay rates. For the $A_0$, similar results (not included here) than those for the $H_0$, are found.

In fig.~\ref{contour} we show the behaviour of the SUSY-QCD contributions to $\Gamma (h_0 \to b \bar s + s \bar b)$ with the different MSSM parameters. We can appreciate the above mentioned grow with $\tan \beta$, the expected decrease with $M_0$, an approximate symmetric behaviour under $\mu \to -\mu$, a growing with $|\mu|$ for moderate $\mu < 600$ GeV, and a nearly independence with the trilinear parameter A.  

\psfrag{M2}{$\,M_2$}
\psfrag{A}{{$A$}}

\begin{figure}
\psfig{figure=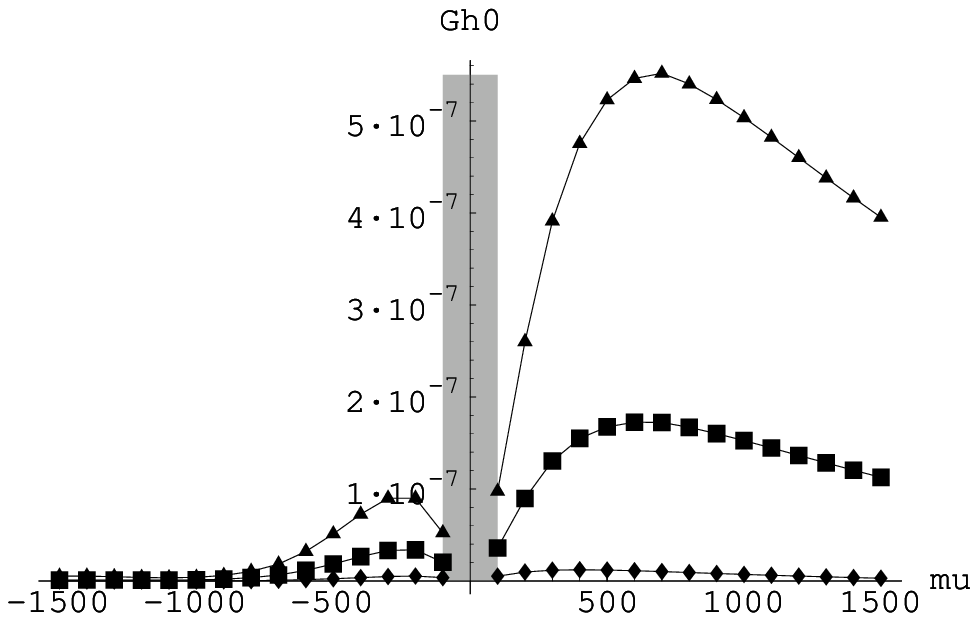,height=1.5in}
\psfig{figure=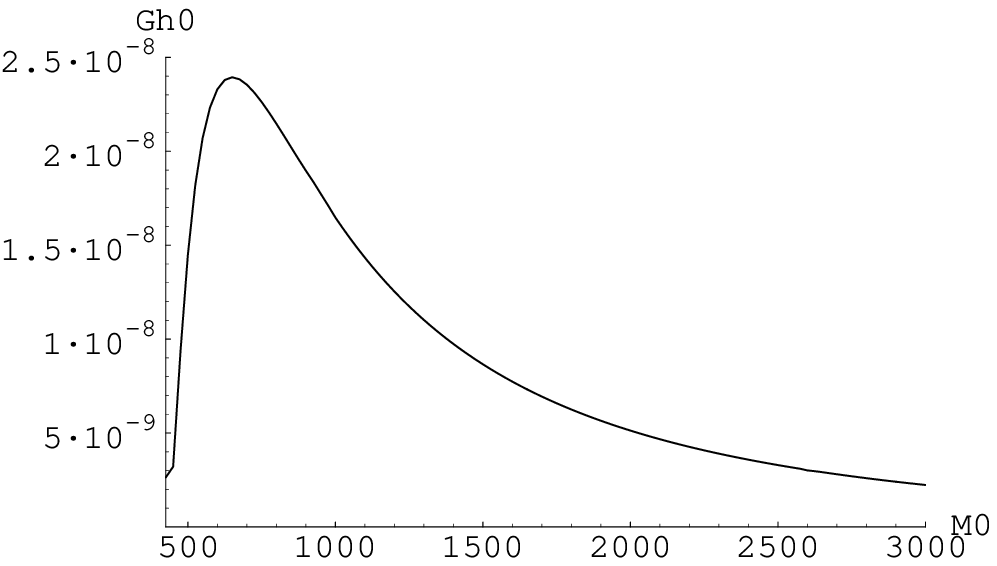,height=1.5in}
\caption{SUSY-EW $\Gamma (h_0 \to b \bar s + s \bar b)$
in GeV as a function of $\mu$ (GeV).
Triangles, boxes, and diamonds correspond 
to $m_A=200, 250, 400$ GeV, respectively (left).$\Gamma (h_0 \to b \bar s + s \bar b)$ as a function of $M_0$ (right). 
\label{hbs_mu}}
\end{figure}

We show in Fig.~\ref{hbs_mu} (left) the 
behaviour of the SUSY-EW contributions to
$\Gamma (h_0 \to b \bar s + s \bar b)$ as a function 
of the $\mu$ parameter for three different values of $m_A$. 
The shaded regions in these figures correspond to the region excluded
by LEP bounds on the chargino mass $|\mu| < 90$ GeV.
The width for the $H^0$ decay (not shown) is approximately symmetric 
under $\mu \to -\mu$, depending of the $m_A$ values, 
while, as we can see, the $\Gamma (h^0 \to b \bar s + s \bar b)$ width 
is more asymmetric with respect to the sign of $\mu$, but all decay widths increase with $|\mu|$ for $|\mu| < 500$ GeV,
then reach a maximum, and finally decrease. Regarding the behaviour at
very small $\mu$ values, we have also found that the widths do not
vanish at $\mu=0$. The origin of this comes entirely from contributions 
driven by electroweak gauge couplings.

Fig.~\ref{hbs_mu} (right) shows the behaviour of the SUSY-EW $h^0$ decay as a function of  $M_0$. 
The $h^0$ decay width has a small value for light $M_0$ due to
the fact that chargino and neutralino contributions have opposite sign
there. For higher values of $M_0$, the neutralino contributions change
sign and the partial cancellation disappears, therefore, the decay width 
increases until it reaches a maximum and then decreases for heavier squarks.
The previously mentioned cancellation for small values of $M_{0}$ is
less obvious for the heavy Higgs, the clearly visible effect is the
decrease due to the growing squark masses, which is slower in the latter case. 

In the following we study the behaviour of the corresponding FCHD with 
respect to $\lambda$. 
Investigating the contributions from gluinos, charginos and neutralinos separately, 
we found that the gluino/neutralino contributions increase monotonically
with $\lambda$, being exactly zero for $\lambda=0$, as expected. 
On the other hand, the contributions from charginos show explicitly the two FC 
sources, CKM and quark-squark misalignment. In fact,
the chargino contribution is different from zero for $\lambda=0$. 
The non-zero value at $\lambda=0$ is due to CKM mixing which is not present
in neutralino (or gluino) loops.  
The CKM effect and the effect of squark mixing compete in the SUSY-EW case in some regions of parameters but for larger values of $\lambda$ the non-CKM flavour-mixing effect
dominates.

\begin{figure}
\psfig{figure=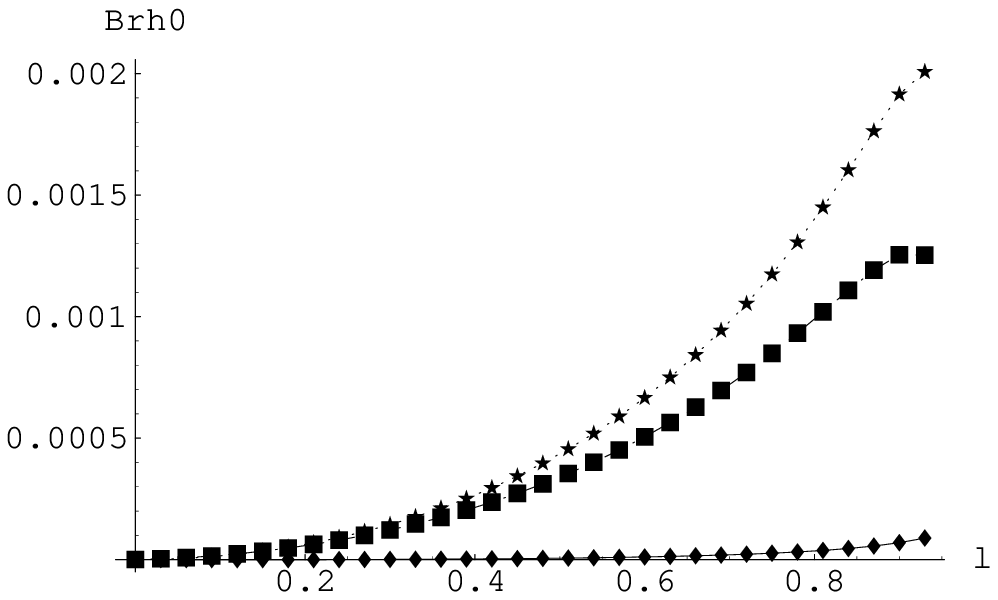,height=1.2in}
\psfig{figure=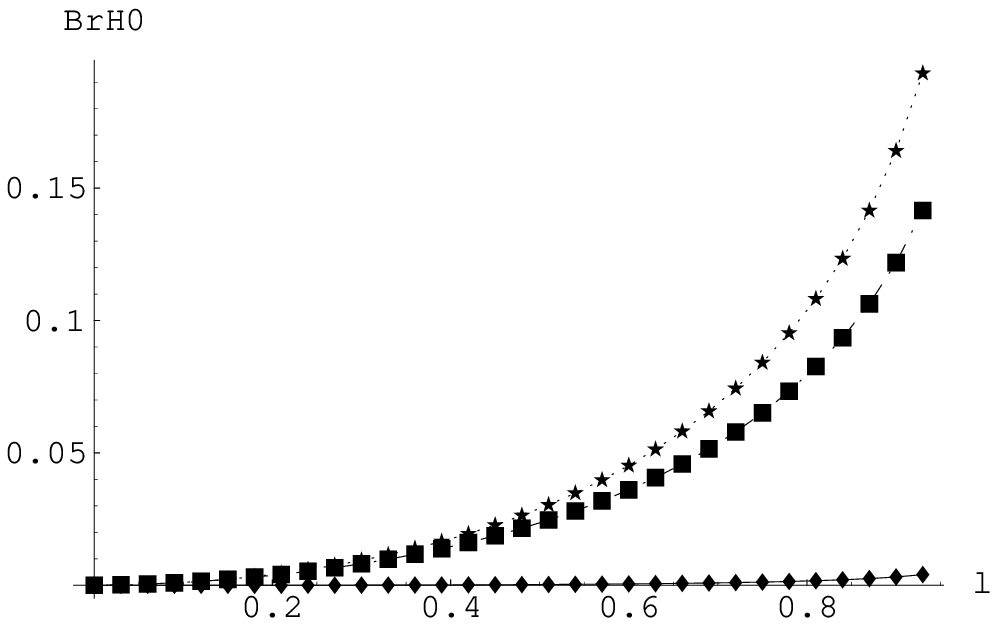,height=1.2in}
\psfig{figure=neubert_plot+error.epsi,height=1.2in}
\caption{SUSY-EW (Diamonds), SUSY-QCD (stars) and total (boxes) contributions
  to the $Br( H_x \to b \bar s + s \bar b)$ $H_x =(h^0, H^0)$ (a) and (b). The prediction for Br($b \to s \gamma$) in the MSSM is shown in (c) as a 
function of $\lambda$ for the chosen MSSM parameters specified in the text (solid line). Short-dashed horizontal lines limit the experimental allowed region and long-dashed ones represent the spread out in the MSSM prediction induced by the theoretical uncertainty in the SM contribution of 99 $\%$ C.L$^5$.
\label{ewglu}}
\end{figure}

In fig.~\ref{ewglu} (a) and (b) we can see the SUSY-QCD, SUSY-EW and total contributions for $m_A=400$ GeV, $\tan \beta =35$, $M_o=800$ GeV, $A=500$ GeV and $M_2=300$ GeV. Note that the absolute value for 
gluino, chargino and neutralino contributions grow with 
$\lambda$ separately and that the decay rates of $h^0$ are 
much larger for smaller values of $m_A$  
and therefore yield larger values of the branching ratio 
$Br ( h^0 \to b \bar s + s \bar b)$. 
Notice also that not all values of $\lambda$ plotted in figs.~\ref{ewglu} are compatible with $b \to s \gamma$ constrains for the chosen values of the MSSM parameters (see fig.~\ref{ewglu} (c)). However, we can still find values of $\lambda$ compatible with $b \to s \gamma$ that still produce large Higgs branching ratios. For instance, the value $\lambda=0.3$ is allowed, and the branching ratio is around $2 \times 10^{-4}$ for $h_0$ and $10^{-2}$ for $H_0$, which are several orders of magnitude larger than the SM value, $Br(H_{SM} \to b \bar s + s \bar b) \sim 4\times 10^{-8}$ for $m_{H_{SM}}= 114$ GeV.
A more exhaustive analysis on the allowed values for the SUSY-QCD corrections of these MSSM Higgs boson branching ratios being restricted from their
 correlation with the radiative B-meson decays ($b \to s \gamma$), can be found in~\cite{sola_last}. 


\psfrag{Brb}{$Br (b \rightarrow s + \gamma)$}

\section{Non-decoupling behaviour of heavy SUSY particles}
One important feature of these observables is their non-decoupling behaviour, which means that the FC effects remain 
non-vanishing even in the most pessimistic scenario of a very heavy 
SUSY spectrum. 
The origin of this non-decoupling behaviour in these contributions
is the fact that the mass suppression induced 
by the heavy-particle propagators is compensated by the mass parameter 
factors coming from the interaction vertices, this being generic in 
Higgs--boson physics. 
The non-decoupling contributions to effective FC Higgs Yukawa couplings to quarks have 
also been studied in the effective-Lagrangian approach~\cite{Demir,dedes}.  We will use here instead
the full diagrammatic approach which has the advantage of taking into account
  all SUSY loop contributions (not just the dominant ones) and is valid for all
$\tan \beta$ values. Since, on the other hand,
 we are not using the mass--insertion approximation, our results are more 
general, being valid for all values of the FC parameter $\lambda$. We 
also checked that our results converge in the large $\tan \beta$ limit and 
for small $\lambda$ values to the mass insertion approximation results of  
the effective Lagrangian approach.


In the following we present the results for the form factors $F_{L,R}$ defined as \\$i F = -i g\,  \bar u_q  
  [F_L^{qq'} (H) P_L + F_R^{qq'} (H) P_R]  v_{q'} H$ in the large $M_{SUSY}$ limit, keeping just the leading $\mathcal{O}(\frac{M_{EW}}{M_{SUSY}})^0$ term in this expansion and where we assume a common $M_{SUSY}$ mass scale,
\begin{eqnarray}
&&\hspace*{-0.5cm}F_{L_{\tilde g}}^{(x)}=
 \frac{\alpha_S}{6 \pi}\frac{m_b}{2 m_W \cos\beta} \left(\sigma_2^{(x)}+\tan\beta\, \sigma_1^{(x)*} \right)F(\lambda) \nn \\
&&\hspace*{-0.5cm}F_{L_{\tilde \chi^{\pm}}}^{(x)}= 
\frac{\alpha_{EW}}{4 \pi} \frac{m_b}{2 m_W \cos\beta}\left[\frac{1}{8 m_W^2 \sin ^2 \beta} \left[\left(V_{{\scriptscriptstyle CKM}}^{tb} V_{{\scriptscriptstyle CKM}}^{cs} m_c^2 +V_{{\scriptscriptstyle CKM}}^{cb} V_{{\scriptscriptstyle CKM}}^{ts} m_t^2 \right)F(\lambda) 
\right.\right. \nn \\
&& \left. 
+ \left(V_{{\scriptscriptstyle CKM}}^{cb} V_{{\scriptscriptstyle CKM}}^{cs} m_c^2 +V_{{\scriptscriptstyle CKM}}^{tb} V_{{\scriptscriptstyle CKM}}^{ts} m_t^2 \right)J(\lambda)\right] -\frac{1}{4} \left[ \left(V_{{\scriptscriptstyle CKM}}^{cb} V_{{\scriptscriptstyle CKM}}^{cs} + V_{{\scriptscriptstyle CKM}}^{tb} V_{{\scriptscriptstyle CKM}}^{ts}\right) J(\lambda)
\right. \nn \\
&& \left. \left. 
+\left(V_{{\scriptscriptstyle CKM}}^{cb} V_{{\scriptscriptstyle CKM}}^{ts} + V_{{\scriptscriptstyle CKM}}^{tb} V_{{\scriptscriptstyle CKM}}^{cs}\right) F(\lambda)\right]\right] \left(\sigma_2^{(x)}+\tan\beta\, \sigma_1^{(x)*} \right)\,,
\label{eq.hbsLLonescalecharginos} \\ \nn \\
&&\hspace*{-0.5cm}F_{L_{\tilde \chi^0}}^{(x)} = 
-\frac{\alpha_{EW}}{4 \pi} \frac{m_b}{2 m_W \cos\beta}
\left[\frac{1}{8}\left(1+\frac{5}{9}\tan^2\theta_W\right)\left(\sigma_2^{(x)}+\tan\beta\, \sigma_1^{(x)*} \right)\right]F(\lambda)\,,
\label{eq.hbsLLonescaleneutralinos} 
\end{eqnarray} 

where, $F(\lambda)=\frac{2}{\lambda^2} [(\lambda +1)\ln(\lambda+1) + 
(\lambda -1)\ln(1-\lambda) - 2\lambda]$, $J(\lambda) = \frac{2}{\lambda^2} [(\lambda +1)\ln(\lambda+1) -
(\lambda -1)\ln(1-\lambda)]$, $\sigma_1^{(x)}=(\sin\alpha,-\cos\alpha,i \sin\beta)$ and $\sigma_2^{(x)}=(\cos\alpha,\sin\alpha,-i\cos\beta)$. The results for $F_R$ are 
like the previous ones but replacing 
$m_b\rightarrow m_s$, $m_c\rightarrow m_t$ and taking the complex
conjugate.

\begin{figure}
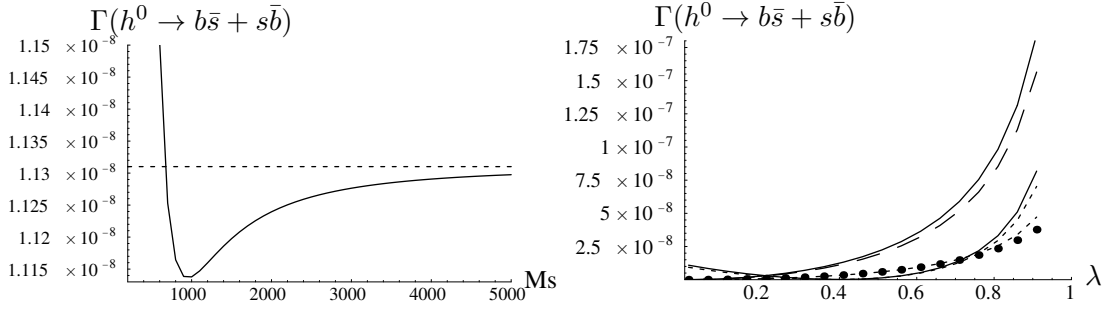

\psfig{figure=anchura_total.h0.epsi,height=1.5in}
\psfig{figure=comparison_lambda_variation.h0.epsi,height=1.5in}
\caption{Non-decoupling behaviour of SUSY-EW $\Gamma(h_0 \to b\bar s + s\bar b)$ in GeV with 
$M_0=\mu=A=M_{\tilde g}=M_S$, $M_1 = \frac{5}{3}\frac{g_1^2}{g_3^2}  M_{\tilde g}$ 
and $M_2 =\frac{g^2}{g_3^2}  M_{\tilde g}$,
  and for $\tan\beta=35$, $\lambda = 0.5$, $m_{A}=m_{H^0}=250$ GeV and
$m_{h^0}=135$ GeV. Exact one-loop results in solid lines and expansions in dashed lines (left). SUSY-EW $\Gamma (h_0 \to b \bar s + s \bar b)$
  as a function of $\lambda$ (right).
\label{widthdecouphbs_Ms}}
\end{figure}

In the following, we show graphically some of the main features. 
For brevity, we choose to show in fig.~\ref{widthdecouphbs_Ms} just the SUSY-EW contributions but the SUSY-QCD ones behave in a similar way.
 We can see that, for large values of $M_S \equiv M_{SUSY}$, the exact partial widths tend to a non-vanishing value, characteristic of the non-decoupling
behaviour. Besides, the exact results are very well described, even for moderate $M_S$ values (say $\geq 600$ GeV), by our previous asymptotic results, which make these short formulas very useful for future phenomenological Higgs boson studies.

The dependence of the total chargino and neutralino contributions    
on the FC parameter $\lambda$ are shown in Fig.~\ref{widthdecouphbs_Ms} (right). The exact one-loop chargino (neutralino) contribution is plotted in 
solid (dotted) lines and the dashed lines correspond to their approximate asymptotic results, given by the previous formulas. The lower lines take into account both possible FC effects while the upper ones correspond to the charginos contribution setting $V_{CKM}=I$. We can appreciate that these effects interfere destructively.
 We can also see that the neutralino and chargino contributions depend strongly on the chosen parameters and can be comparable in size for a large range in $\lambda$, therefore, neglecting the neutralino contribution, as usually done in the literature, is not a good approximation.  

Regarding the so-called decoupling limit where $m_A>>m_{EW}$, we see clearly that the SUSY loop contributions, in the $h_0$ case, go to zero, recovering, as expected, the SM result.

\section{Conclusions}
We have computed the genuine SUSY one-loop quantum effects to 
flavour-changing MSSM Higgs-boson decays into second and third generation quarks using
 the full diagrammatic approach and therefore our results are valid for all
 $\tan \beta$ values and for all values of the 
flavour-mixing parameter $\lambda$. We find that for moderate $\tan \beta$ 
and $\lambda$ values, some of the contributions usually neglected 
in the effective--Lagrangian approach can be sizable and, therefore, a 
realistic estimate of the branching ratios should rely better on the
full diagrammatic approach. 
 After  
analyzing in full detail the dependence of the FCHD partial widths, with all 
the relevant MSSM parameters and $\lambda$, we found large rates and these are very 
sensitive to $\tan \beta$, $\mu$, and $\lambda$. The branching ratios 
grow with both $\tan \beta$ and $\lambda$ and reach quite sizable values 
in comparison with the SM ones, in the large $\tan \beta$ and $\lambda$ 
region. For instance, for the following choice allowed by the $b\to s \gamma$ constrains, $\lambda=0.3$, $\tan \beta =35$, $m_A = 400$, $M_o=800$ GeV, $A=500$ GeV and $M_2=300$ GeV, we found branching ratios of $\mathcal{O} (10^{-4})$ for
 the $h^0$ and $\mathcal{O} (10^{-2})$ for the $A^0$ and $H^0$, which are both some orders of magnitude larger than $Br_{SM} \approx \mathcal{O} (10^{-8})$.
As expected, the SUSY-EW 
contributions are subdominant with respect to the SUSY-QCD ones,
but they contribute with opposite sign and important interference
effects, which modify the SUSY-QCD effects remarkably, appear.

An interesting feature of these radiative contributions is their 
non-decoupling behaviour for large values of the SUSY particle masses, {\it i.e.}, even in the most pessimistic scenario of very large SUSY mass parameters, FC effects remains and they can be sizable (a set of analytical asymptotic results 
can be found in refs.~\cite{Maria,Maria2}).

In conclusion, the results presented in this talk indicate that 
FCHD constitute an interesting scenario for indirect searches of supersymmetry, with important contributions in some regions of the MSSM parameter space, that remain even for a very heavy SUSY spectra.
 
\section*{Acknowledgments}
We would like to thank W.~Hollik, F.~Merz, S.~Pe\~naranda and D.~Temes for fruitful collaborations, which have been the basis for this talk. A.~M.~C. wishes to thank the organizers for inviting her to a very enjoyable meeting and L.~Silvestrini for interesting comments.

\end{document}